\documentclass{article}
\usepackage{spconf,amsmath,graphicx}

\title{\LARGE Whisper-GPT - \\
Continuous Discrete Hybrid Representation \\Language Models For Speech And Music}
\name{Prateek Verma}
\address{Stanford University, Stanford CA, 94305}
\begin{document}
\maketitle
\begin{abstract}
We propose \textbf{WHISPER-GPT}: A generative large language model (LLM) for speech and music that allows us to work with continuous audio representations and discrete tokens simultaneously as part of a single architecture. There has been a huge surge in generative audio, speech, and music models that utilize discrete audio tokens derived from neural compression algorithms, e.g. ENCODEC. However, one of the major drawbacks of this approach is handling the context length. It blows up for high-fidelity generative architecture if one has to account for all the audio contents at various frequencies for the next token prediction. By combining continuous audio representation like the spectrogram and discrete acoustic tokens, we retain the best of both worlds: Have all the information needed from the audio at a specific time instance in a single token, yet allow LLM to predict the future token to allow for sampling and other benefits discrete space provides. We show how our architecture improves the perplexity and negative log-likelihood scores for the next token prediction compared to a token-based LLM for speech and music. \footnote{This work was proto-typed using the support of the Stanford Institute of Human-Centered AI (HAI) through a Google Cloud grant for the academic year 2023-2024. The authors thank Google and Stanford HAI for this initiative. }

\end{abstract}
\section{Introduction and Related Work}
\label{sec:intro}
Transformers and LLMs have exponentially surged the advancements in artificial intelligence across fields and modalities. Initially designed for machine translation \cite{vaswani2017attention}, they were quickly adopted for a variety of fields such as text \cite{brown2020language}, raw audio waveforms\cite{verma2021generative,verma2022goodbye}, acoustic and music tokens \cite{huang2018music, verma2020framework,borsos2023audiolm}, videos\cite{yan2021videogpt} to name a few. It quickly superseded widely popular convolutional architectures too for audio \cite{verma2021audio,gong2021psla} and vision \cite{dosovitskiy2020image} perception. With the recent Gemini family combining modalities from the ground up \cite{team2023gemini} or multi-modal models like Chameleon, the future will bring in even more exciting advancements toward AGI that will allow architectures to see, hear, read, and reason like humans. Almost all papers in speech and music adopt these architectural advances by posing audio and music language modelling as generative models on acoustic tokens. The seminal paper in this direction was VQ-VAE architecture \cite{oord2017neural}, which learned discrete representation from continuous audio and modelled these discrete representations using neural architectures. This was further extended to music by OpenAI in famous JukeBox architecture\cite{dhariwal2020jukebox}, which utilized multiscale discrete representations conditioned on lyrics and text. One of this method's main drawbacks was using WavNet\cite{oord2016wavenet} based vocoders that led to the slow generation process. This led to early works on using codec-based methods; audio compression algorithms were trained on audio waveforms in an autoencoder-like setup in works like Soundstream or ENCODEC \cite{zeghidour2021soundstream,defossez2022high} to get discrete tokens. This recipe has become defacto in speech and music generative architectures such as Textless NLP\cite{lakhotia2021generative,kharitonov2022textless}, AudioLM\cite{borsos2023audiolm}, MusicLM\cite{agostinelli2023musiclm}, MusicGen\cite{copet2024simple} to name a few. Like VALL-E\cite{wang2023neural}, all of them first model coarse Sound-stream or ENCODEC tokens followed by fine-grained tokens. Our work also draws inspiration from using hybrid input representations with early fusion for solving downstream audio tasks \cite{verma2020deep,wang2021multi}. This paper will tackle generating only the coarsest tokens for all experiments, as other tokens are generated and conditioned on them in a causal/noncausal manner in AudioLM/VALL-E. We also draw inspiration from Whisper, which is a seq-to-seq encoder-decoder model. It goes from mel-spectrogram input to GPT tokens for ASR. Whisper combines continuous input-discrete output in noncausal setups. This work asks -- 

\textbf{Can we devise an architecture that utilizes both continuous and discrete representation in the LLM setup?} 

The paper's contributions are as follows: 1. To the best of our knowledge; we introduce the first hybrid generative causal architecture audio shown for speech and music that combines continuous acoustic representation like mel-spectrogram with discrete acoustic tokens. 2. We adapt a Whisper-like architecture, a noncausal ASR seq, to a seq architecture for generative modelling. We replace the Whisper encoder with a decoder and, on learned representation, carry early fusion with decoder-only architecture operating on acoustic tokens. 3. We showcase significant improvements in the next token prediction for the music and speech dataset in a VALL-E-like setup, which predicts the coarsest acoustic token with the hybrid.

\begin{figure*}[t]
  \centering
  \hspace*{8.8pt}
  \includegraphics[width=0.75\linewidth,height=4.8cm]{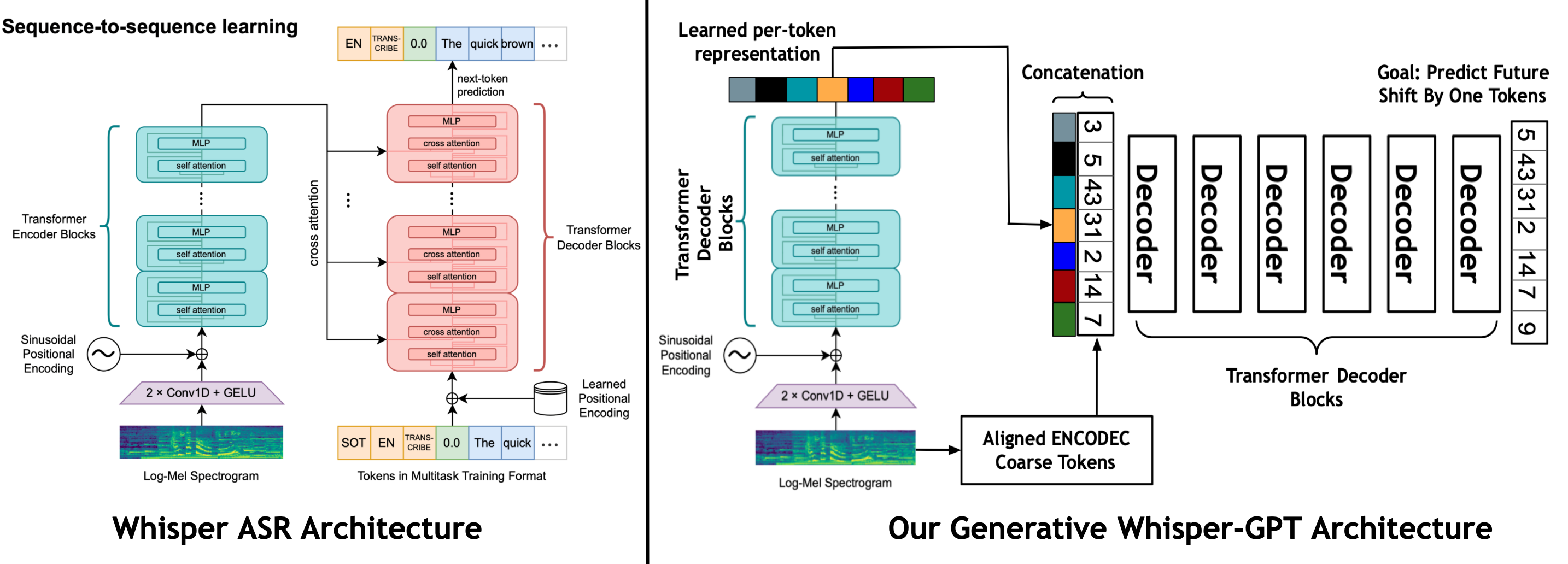}
  \caption{ (Left) Whisper Architecture proposed by OpenAI \cite{radford2023robust} which treats ASR as a sequence to sequence which takes in mel-spectrogram slices and decodes it token by token. It has a Transformer Encoder stack on the spectrogram followed by a Transformer decoder, trained for the shift-by-one token prediction, and the cross-attention module on learned spectrogram representation. (Right) Our generative model combines both continuous and discrete representations. We align the spectrogram and ENCODEC coarse tokens. Instead of a Transformer encoder, we pass spectrogram slices through lightweight decoder blocks. The learned representation per-token slice is concatenated with discrete tokens corresponding to the spectrogram slice to have a decoder Transformer stack, trained on shift by one next token prediction, similar to a typical LLM pre-training.}
  \label{fig:speech_production}
\end{figure*} 
\section{Dataset}\label{sec: data}We report results on two domains: music and speech. For the case of speech, we use the LibriSpeech TTS dataset \cite{zen2019libritts}. We chose this over LibriSpeech as it contains material derived from the original LibriSpeech corpus for TTS research. It also removes utterances with significant background noise. Finally, it only consists of 24 kHz sampled utterances and removes other lower sampling rates, thereby matching the rate for ENCODEC \cite{defossez2022high} acoustic tokenizer rate. For the case of music, we use publically available music recordings of instrumental music, namely piano, saxophone, harp, flute, violin, marimba, etc., for a total of 200 hours of music. We extract 64-channel mel-spectrogram input to the hybrid models and adjust the hop/window length to match the 75Hz rate of ENCODEC acoustic tokens. We extract causal acoustic tokens at the coarsest level for the entire paper, similar to as described in VALL-E \cite{wang2023neural} and predict the next tokens i) using discrete token-based GPT architecture ii) a hybrid mel-spectrogram-based architecture that takes in both the current mel-spectrogram slice and coarsest discrete acoustic token.
\section{Methodology}
\label{sec:format}
All the models discussed in the paper are Transformer decoder-only architecture. The recipe has become typical for music \cite{copet2024simple} and speech generation \cite{lakhotia2021generative}, where the coarsest tokens are first predicted using auto-regressive architecture followed by finer acoustic tokens, conditioned on the coarsest tokens using auto/non-auto regressive architecture \cite{borsos2023audiolm}, \cite{wang2023neural}. This is primarily done as it is challenging to model long-context acoustic tokens for audio. This work aims to determine how well we can model the coarsest tokens, as errors in modelling the coarser tokens will lead to the finer tokens being modelled incorrectly as they are conditioned on the coarsest token, thus affecting model performance. We report how well the model performs in pre-training instead of post-training, as we are interested in building better foundational architectures in this paper. The other motivation is to push the capabilities of smaller transformer-based decoder architectures that can be trained in academia. The current architectures, such as those used in AudioLM and Vall-E, trained from scratch, are beyond the reach of most academic setups. We do not use other techniques such as distillation \cite{hinton2015distilling} or pruning \cite{sun2023simple} that often uses the capabilities of a larger architecture to push the performance of a smaller architecture. Instead, we use better input representation to achieve the performance of a much larger architecture, which is a purely token-based model. To our knowledge, we propose the first hybrid large language model that utilizes continuous representations like mel-spectrogram and acoustic tokens in a causal setup. Given the extended context, using a pure discrete token-based architecture is not easy to model; for ENCODEC, the input audio would be represented by 75x8 = 600 tokens per second for coarser to finer tokens. For all the experiments proposed in our paper, we use a context length of 10s. Modelling 6000 tokens would have exponentially increased the training times of our architecture: such large context lengths are difficult to handle using attention-based methods in LLM setup. Further, building a continuous audio foundational architecture based on mel-spectrogram has its challenges, and sampling from a continuous space to bring diversity in generated output is difficult. Finally, we would have relied on Vocoders or algorithms like Griffin-Lim to return to the original audio. By combining a continuous audio representation like mel-spectrogram and discrete acoustic tokens, we retain the best of both worlds: Have all the information needed from the audio at a specific time instance in a single token, yet allow the LLM to predict the future token to allow for sampling and other benefits discrete space provides.\begin{figure}[t]
  \centering
  \includegraphics[width=0.85\linewidth]{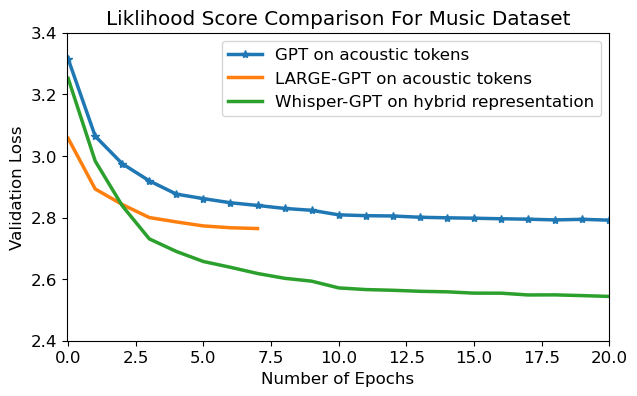}
  \caption{Comparison of GPT on coarse acoustic tokens with i) GPT-L ii) Our hybrid continuous-discrete representation.}
  \label{fig:speech_production}
\end{figure} 
\vspace{-0.85cm}
\subsection{Baseline Architecture}
\vspace{-0.2cm}All the architectures proposed in Transformer decoder architecture are widely used in \cite{wang2023neural}. We keep the architecture topologies the same, with the same number of parameters for speech and music, as they both are based on ENCODEC. It is one of the most commonly used open-source tokenizers for speech and music synthesis since its release in 2022. The baseline architecture consists of 8 Transformer Decoder layers with eight attention heads, an embedding dimension of 64, with the size of the feed-forward dimension four times that of the embedding dimension typically used in Transformer recipes. The final layer of the Transformer is then piped to a dense layer of 2048 neurons followed by a 1024 dim dense layer equal to the vocabulary of the ENCODEC. This has the total number of parameters as 3.7 million. All models were trained for 25 epochs on random crops of 750 token context length from LibriSpeech and Music dataset using Adam optimizer. The initial learning rate was chosen to be 2e-4 and decayed to 1e-4 after ten epochs. As with LLM pre-training, the goal is to predict the next token given the past context while minimizing the cross-entropy loss with ground truth tokens.

\subsection{Scaled GPT Architecture} We compared our baseline architecture with a medium-sized GPT-2 architecture. Both of our architectures, i.e., the baseline and GPT-L architecture, are the same as that of VALL-E \cite{wang2023neural} except that it is a shrunk-down version of it in terms of model dimension, number of layers, and number of heads. For reference, VALL-E, trained on acoustic tokens precisely the same way we have described, has a model dimension of 1024 with the number of layers as 12 and 16 attention heads with a total of around 900 million parameters. This is far beyond anything that can be trained from scratch in an academic setup. Hence, we first proposed a shrunk-down baseline architecture. We then trained the largest possible GPT architecture that we could train using the resources available at our disposal. We used 16 attention heads with eight layers and a model dimension 256, resulting in 40 million parameters. This is our scaled GPT architecture, which we call GPT-Large, ten times larger than our baseline model. We compare our hybrid architecture and the baseline architecture with this model. GPT-L is also trained with the same recipe for 25 epochs. For both of these architectures, we trained them on 10 seconds of context or 750 tokens as context length. 
\subsection{Hybrid Architecture} For our proposed hybrid architecture, we retain the GPT branch as described in the baseline section. However, we have a Whisper-inspired Transformer Decoder stack that fed the mel-spectrogram of the original input audio from which we extracted ENCODEC tokens. We first compute the log-mel spectrogram with hop length to match the ENCODEC sampling frequency, i.e., 75 Hz. We computed 64 mel spectrogram bins with 40ms window length and converted the sampling frequency to match the sampling rate of ENCODEC tokens, i.e., 24 KHz. This gave us a representation of  64 mel-bins x 75 for every second of audio. Even though ENCODEC tokens are fully causal, further care was carried out to maintain causality by ensuring that the window computation is causal, i.e., not centred for every hop to prevent leakage. In addition, instead of predicting the next token from the spectrogram slice, i.e., the default shift by one prediction, we opted for the shift by 2, i.e., the current ENCODEC coarse token is predicted from all past slices, barring the most recent two slices. We retain the same architecture as the Whisper-encoder with two modifications: i) We shrink it in terms of a number of parameters, and ii) We have causal attention masks so as not to attend to future spectrogram slices to predict the current ENCODEC token.
For the Whisper decoder, we have six layers, with the model dimension of 32 (half as our GPT baseline) and the feed-forward dimension four times that of the model dimension. We pass single spectrogram slices as input and normalize them across the corpus to have a unit Gaussian distribution. In this case, we use a single linear layer of 2048 neurons followed by a dense layer to map it to the embedding dimension of the model (32). Relative sinusoidal positional embeddings are added, followed by a stack of decoder layers. We call it the Whisper-Decoder, which resembles the Whisper-Encoder part of the Whisper architecture. The output of the last layer of Whisper-Decoder has a dimension of 32 for all the tokens present in the data, i.e., 750 for context length chosen as 10s. We now use an embedding dimension of 32 for the input embedding matrix for ENCODEC acoustic tokens and add sinusoidal positional embeddings, as shown in Figure 1. The output of the input embedding matrix is then concatenated with the last layer of the Whisper-Decoder output, which has a dimension of 32 for all tokens. This gives us a 64-dimension vector for all of the inputs (750) of the present, i.e., 32 coming from the input embedding matrix of the ENCODEC tokens and 32 coming out of the last layer of the Whisper-Decoder block. This concatenated 64-dimensional vector is then passed to the stack of decoder layers, which are the same as our baseline, i.e., eight layers with model dimension 64 and feed-forward dimension four times that of the embedding dimension and eight attention heads. The output of the last layer is then passed to a dense layer of 2048 neurons, followed by a 1024-dimensional dense layer to conform with the vocabulary of the ENCODEC tokens. This has the total number of parameters as 4.1 million. All models were trained for 25 epochs on random crops of 750 token context length from LibriSpeech and Music dataset using Adam optimizer. The initial learning rate was chosen to be 2e-4 and decayed to 1e-4 after ten epochs. The goal is to predict the next token given the past context while minimizing the cross-entropy loss. 
\begin{figure}[t]
  \centering
  \includegraphics[width=\linewidth]{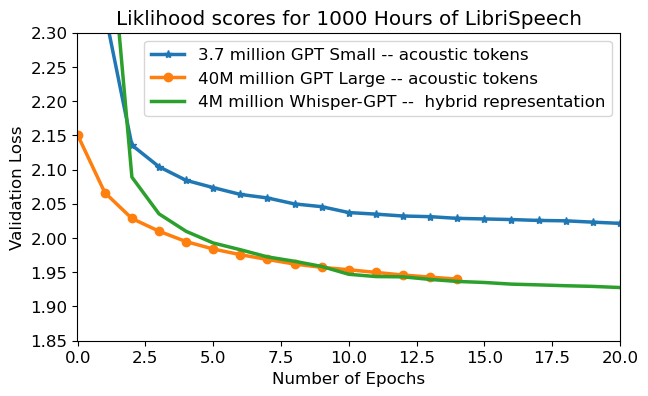}
  \caption{Comparison of GPT on coarse acoustic tokens with i) GPT-L ii) Our hybrid continuous-discrete representation.}
  \label{fig:speech_production}
\end{figure} 
\section{Results and Discussion}
\label{sec:results}
We carry out various ablation studies to report how well our generative model does in the following section. We are only interested in quantifying the likelihood scores for the generative architecture pre-training. The rationale is as follows - i) We propose a hybrid GPT-like architecture. Given the previous context, these architectures are trained on a simple objective to predict the next token, typical of LLM pre-training. In order to improve on this, models are usually scaled in the number of parameters with identical topologies. We report our results on a baseline GPT-S architecture and compare it with scaling it to an architecture ten times in size. We then compare our results with our hybrid architecture, as reported in Tables 1 and 2. We see that for both speech and music, we outperform an architecture that is ten times larger than the baseline model, thus pushing the boundaries of large language models by using a better input representation. By combining continuous audio representation like spectrogram and discrete acoustic tokens, we retain the best of both worlds: Have all the information needed from the audio at a specific time instance in a single token, yet allow LLM to predict the future token to allow for sampling and other benefits discrete space provides.\vspace{-0.3cm}\begin{table}[ht]
  \caption{Negative-log likelihood (NLL) and perplexity scores for our proposed hybrid architecture, baseline GPT-Small and a GPT-Large 10x larger than GPT-Small for LibriSpeech.}
	\centering
	\begin{tabular}{|c|c|c|c|c|c|}
		\hline
		Model & \# of Param & NLL  & Accuracy & PPL \\\hline
        Baseline GPT-S  & 3.7 million  &  2.02 & 34.18\% & 7.54 \\
		GPT-L  & 40 million  & 1.94  &34.82\%  & 6.96 \\
        \textbf{Hyrbid LLM}  & \textbf{4 million}   & \textbf{1.93}  & \textbf{35.05\%} & \textbf{6.96} \\\hline
	\end{tabular}
	\label{tab:example}
\end{table} One of the reasons we see our hybrid architecture outperforms significantly better for music tokens vs. speech tokens is that music has much more information present about instrumentation, pitch, timbre, and multiple harmonic components, which is hard to model as compared to LibriSpeech which is a single speaker saying English words. Further, for music signals, just modelling the coarsest token might not capture all the variation as compared to having a hybrid-like architecture that looks at both the acoustic token representation and the mel-spectrogram-like representation. We utilize hybrid representation with fewer parameters. 
\vspace {-0.5cm}
\begin{table}[ht]
  \caption{Negative-log likelihood (NLL) and Perplexity (PPL) scores for our proposed hybrid architecture, baseline GPT-Small and a GPT-Large 10x larger than GPT-Small for Music}
  
	\centering
	\begin{tabular}{|c|c|c|c|c|c|}
		\hline
		Model & \# of Param & NLL  & Accuracy & PPL \\\hline
        Baseline GPT-S  & 3.7 million  & 2.78  & 34.96\% & 16.12 \\
		GPT-L  & 40 million  &  2.77 & 35.72\%  & 15.96 \\
        \textbf{Hyrbid LLM} & \textbf{4 million}   & \textbf{2.52} & \textbf{38.47\%} & \textbf{12.43} \\\hline
		
	\end{tabular}
	\label{tab:example}
\end{table}
\vspace{-0.3cm}
\section{Conclusion And Future Work}
\label{sec:results}
We show how one can build hybrid generative architectures for speech and music using hybrid continuous-discrete representation. The proposed architecture outperforms a purely token-based model by combining continuous audio and discrete token-based representations. Our experiments show that by using hybrid continuous-discrete representation for two datasets, we can match the performance of using acoustic tokens of a 40M parameter architecture with a model with only 4M parameters. This retains the best of both worlds: allowing the model to have complete information on the audio signal in a single token embedding while still predicting discrete acoustic tokens that we can sample from to introduce diversity and variations in the generated output.
\bibliographystyle{IEEEbib}
\bibliography{refs}
\end{document}